\documentstyle[twoside,fleqn,npb,epsfig]{article}
\newcommand{\AmS}{{\protect\the\textfont2
  A\kern-.1667em\lower.5ex\hbox{M}\kern-.125emS}}

\title{Radiative corrections in a minimal extension of the standard
    model.}

\author{J. J. van der Bij\address{Fakult\"at f\"ur Mathematik und Physik,
        Physikalisches Institut,
        Universit\"at Freiburg,\\ H. Herderstr. 3, 79104 Freiburg i.B.,
        Germany }%
        \thanks{This work was supported by the European Union
          under contract HPRN-CT-2000-00149 and by the 
         DFG-Forschergruppe: Quantenfeldtheorie, Computeralgebra und
   Monte-Carlo-Simulation.}
        }

\begin{document}

\begin{abstract}
Radiative corrections are studied within an extension of the
standard model, containing extra singlet scalars. The calculations
determine the effect of a large width of the Higgs boson on
radiative corrections. They throw some light  on the treatment
of unstable particles inside loop-graphs.
\end{abstract}

\maketitle

\section{INTRODUCTION}

The
standard model of elementary particle physics based on the
gauge group $SU(3) \times SU(2) \times U(1)$ is essentially confirmed by
the precision tests at LEP \cite{sriem}. However the Higgs particle 
has not been
found and  a naive analysis of the data gives a bound of
 $m_H > 112 GeV$. However the data are not without trouble. There is
a large difference between the leptonic data, including the 
measured W-mass and the results from the b-physics data at LEP.
The leptonic data by themselves would imply a lower mass for the Higgs
boson than is acceptable from the direct search. While it is
probably premature to conclude that the standard model is definitely
ruled out, the discrepancy points out the importance of studying
alternatives to the minimal Higgs sector of the standard model.

Because the effect is small and the general agreement with the standard
model is there, a natural option is to study the effects of changes
in the Higgs sector alone. The simplest such extension is to
include extra Higgs singlet scalar particles in the model.
Singlets do not couple to ordinary matter directly in a 
renormalizable way, but only to the Higgs sector. Therefore indirect 
effects of such particles appear only at the two-loop level
in precision tests of the standard model. Since effects at the two-loop
level are generically suppressed, significant contributions
of the singlets can only arise if they have at least moderately
strong interactions. When this is the case one can in principle
not limit oneself to two-loop graphs only, but needs to resum an
infinite number of graphs, thereby being forced
 to confront the non-convergent
nature of perturbation theory. This problem is not unique to
Higgs physics, but arises always when one has instable particles
inside loops, since the propagator of an instable particle implies
a Dyson-resummation of the one-particle irreducible contribution to
the two-point function. In gauge theories this leads to complications
regarding gauge invariance and questions regarding high precision
predictions for instance for the process  $e^+e^- \rightarrow  W^+W^-$.

In order to address such questions it is advantageous to have a simple
model, that avoids some of the complications. Also the presence of a
non-perturbative expansion parameter is desirable. These arguments 
lead one to consider the standard model coupled to an O(N)-symmetric
sigma-model\cite{binoth}. By a suitable rescaling of coupling constants 
one
can use the parameter (1/N) as a non-perturbative expansion parameter.
It is this model that will be used in the following. Besides the fact 
that it is essentially the simplest extension of the standard model
and therefore should be studied, the model has some interesting features.
If the extra scalars are light enough, the Higgs boson will decay
into them, giving a large invisible width to the Higgs boson. This
will eliminate practically any signal at the LHC. However 
TESLA\cite{tesla} will
have no problem discovering such a Higgs boson. The decay
products of the Higgs boson are stable and weakly interacting with
ordinary matter. They form a suitable candidate for the dark matter 
of the universe\cite{cosmo1}\cite{cosmo2}.

\section{MNMSM or STEALTH MODEL}

The model of a Higgs sector, containing the standard model
Higgs boson plus an O(N)-symmetric sigma model,
 is essentially
the simplest extension of the standard model and one can call
it the M(inimal) N(on)-M(inimal) S(tandard) M(odel), filling
in the gap between  SM, MSSM and  NMSSM. Because of the hidden nature
of the Higgs boson one can also call it the stealth model.

The Lagrangian density is the following:

\begin{eqnarray} \label{model}
L_{Scalar} &=& L_{Higgs} + L_{Phion} + L_{Interaction}   \nonumber  \\
           & & \hspace{0cm} \mbox{where}\nonumber \\
L_{Higgs}  &=& 
 - \partial_{\mu}\phi^+ \partial^{\mu}\phi -\lambda \,
 (\phi^+\phi - \frac{v^2}{2})^2 \nonumber \\
L_{Phion}  &=& - \frac{1}{2}\,\partial_{\mu} \vec\varphi \, 
\partial^{\mu}\vec\varphi
     -\frac{1}{2} m_{P}^2 \,\vec\varphi^2 - \frac{\kappa}{8N} \, 
     (\vec\varphi^2 )^2 \nonumber \\
L_{Inter.} &=& -\frac{\omega}{2\sqrt{N}}\, \, \vec\varphi^2 \,\phi^+\phi 
\end{eqnarray}  
Here we use a metric with signature $(-+++)$. 
$\phi=(\sigma+v+i\pi_1,\pi_2+i\pi_3)/\sqrt{2}$ is the complex Higgs doublet of 
the SM with the
vacuum expectation value $<0|\phi|0> = (v/\sqrt{2},0)$, $v=246$ GeV. Here, 
$\sigma$ is the physical  
Higgs boson and $\pi_{i=1,2,3}$ are the three Goldstone bosons. 
$\vec\varphi = (\varphi_1,\dots,\varphi_N)$ is a real vector with 
$<0|\vec\varphi|0>= \vec 0$. 
We consider the case, where the $O(N)$ symmetry stays unbroken,
because we want to concentrate on the effects of a finite width
of the Higgs particle. Breaking the $O(N)$ symmetry would lead
to more than one Higgs particle, through mixing. Since the goal of this
exposition is to clarify the role of a finite width
of the Higgs boson inside loop graphs, the model will be analyzed
in the limit, where the phion mass is zero. Also the phion
selfinteractions will be ignored. In this limit fully analytic results
can be presented. A detailed analysis of the general case, which
necessarily becomes numerical, is given in \cite{akhoury}.
The physics of the model in the above limit is quite simple.
One has a Higgs boson that has a larger than usual width,
but there are no other direct physics effects present. Within the
$1/N$ expansion only bubble-graphs containing phions contribute
to the Higgs propagator, which enters the electroweak radiative
corrections.

\section{THE $\rho$ PARAMETER}
\subsection{Definition}

Higgs mass dependent corrections to low energy electroweak parameters,
appear in a number of places. Because the Yukawa couplings
are very small, the Higgs dependent effects are limited to
corrections to the vector-boson propagators, which can be
taken to be the $\rho$ parameter, the mass shift of the 
Z boson and the mass shift of the W boson, often parametrized
by S,T,U or $\epsilon_1, \epsilon_2, \epsilon_3$ \cite{PDR}.
The behaviour for the different quantities is very similar.
If the Higgs mass is much larger than the Z-boson mass,
the effects grow like $log(m_H^2/M_W^2)$. Already for
the present limit $m_H > 112 GeV$, the large Higgs mass limit
is a reasonable approximation. As the behaviour for 
the different quantities is similar, it is sufficient 
to focus on one of them. We will here choose the
so-called $\rho$ parameter. It is the ratio
of neutral to charged current strengths.
The definition is:
\begin{equation}
\rho = G_F^0/G_F^+
\end{equation} 
At the tree level $\rho = 1$. This is peculiar for a
 doublet Higgs field, the reason being the $O(4)$
symmetry of the Higgs potential, which is larger
than the $SU(2) \times U(1)$ symmetry of the standard model as
as whole. This larger symmetry is broken by the 
hypercharge coupling, leading to effects vanishing
with $tg(\theta_W)$.
The correction is given by:
\begin{equation}
\delta \rho = \rho -1 = 
(\delta M_W^2 - cos^2(\theta_W) \delta M_Z^2)/M_W^2
\end{equation}

where $\delta M_W^2$ and $\delta M_Z^2$ are the corrections
to the vector boson masses at $k^2 = 0$.

\subsection{ $\delta \rho$ in the standard model}

Within the standard model the Higgs mass dependent
contribution can be most easily described in the
unitary gauge, where only one-graph contributes,
namely the virtual splitting and recombining of a
vectorboson into a Higgs boson and a vectorboson.
This holds for all the low energy precision variables,
which explains why the contributions are so similar.
This contribution, coming from the Higgs alone is
still divergent, the divergence being canceled by graphs
containing only virtual vectorbosons and photons.
The exact expression becomes quite complicated.
For the Higgs mass dependent contribution it is
therefore advantageous to consider the difference 
with the hypothetical case of $m_H = 0$.
In the limit, that the Higgs mass becomes large
one then finds the following simple result:
\begin{eqnarray}
\delta \rho (SM,m_H) - \delta \rho (SM,m_H=0) =
-\frac{3g^2}{64 \pi^2}    \nonumber \\
( tg^2(\theta_W) log(m_H^2/m_Z^2) + log(1+tg^2(\theta_W))
\end{eqnarray}
  
The last term comes from the fact that one compares diagrams
with different masses inside the loops; it is actually
$log(m_Z^2/m_W^2)$. If one is interested in just the 
leading logarithm, without the constant, one can make an expansion
directly inside the graphs and ignore the masses of the
vectorbosons. One finds:
\begin{eqnarray}
\delta \rho = \frac{g^2}{(2\pi^4)i} tg^2(\theta_W)
(1-1/n) \nonumber   \\
 \int d^nk (k^2)^{-1} (k^2 + m_H^2)^{-1}
\end{eqnarray}
This gives the leading Higgs mass behaviour. The integral
is still divergent, but this divergence is canceled
by the explicit vectorboson graphs.  The extra factor $(k^2)^{-1}$
simplifies the integral, which is useful in particular when more
loops are present.

\section{FINITE WIDTH EFFECTS}
\subsection{Perturbation theory}
We are now ready to calculate higher order effects
in the model coming from the extra interactions
of the Higgs boson to the phions. Within the $1/N$ expansion
the only contributions that arise are graphs with phion
bubbles inserted in the Higgs propagator. The calculations can therefore
be performed in the same style as the one-loop calculations.
Contributions to the $\rho$ parameter at the n-phion bubble level
are therefore determined by integrals of the form:
\begin{equation}
\int d^4k \frac{1}{k^2} \frac{1}{(k^2+m_H^2)^{n+1}}
log^n(k^2/m_H^2)
\end{equation} 

The logarithms in this integral come from the phion-bubbles,
that are easily calculated, since they are just the massless
one-loop propagator graphs. The one-loop renormalization of the 
Higgs mass is taken into account, through the scale choice 
of $\mu = m_H$ in the bubble-graphs.
Going to polar coordinates and using the simplification
due to the factor $(k^2)^{-1}$ it is
now relatively easy to calculate the diagrams.
The result is: 

\begin{equation} 
\delta \rho = \frac {3 g^2 tg^2(\theta_W)}{64 \pi^2}
\sum_{n=1}^{\infty} \int_{0}^{\infty} (-\Delta)^n 
 \frac {log^n(s)} {(s+1)^{n+1}} ds 
\end{equation}
where:
\begin{equation}
\Delta = \frac {\omega^2 v^2}{32 \pi^2 m_H^2}
\end{equation} 
Explicit evaluation of the integrals  gives:

\begin{equation}
\delta \rho = 
 \frac {3 g^2 tg^2(\theta_W)}{64 \pi^2}
\sum_{n=2}^{\infty}  \frac {(-2\pi \Delta)^n} {n} 
|B_n| ( 1 - 2^{1-n})
\end{equation}
where $B_n$ are the Bernouilli numbers.
Since the odd Bernouilli numbers are 0, only the graphs with
an even number of phion-bubbles contribute.
This at least partly explains the observation \cite{veltman},
that at the two-loop level within the standard model
the correction due to the Higgs particle is small.
It was noticed, that the correction is small, not only
because of the loop-factors, but also because of extra cancellations
between different terms in the final result. The above result explains 
that there is no enhancement proportional to the number of
Goldstone bosons, three in the standard model.

\subsection{Borel summation}
The series that we found for the $\rho$ parameter is
obviously divergent. The question is therefore in how far
one can make sense of the series.
A traditional way to look at such a series is to make a Borel
transformation.
The Borel sum is defined as follows.
Given the above series
\begin{equation}
\sum_{n=2}^{\infty} a_n \Delta^n
\end{equation}
we form the new "Borel" series 
\begin{equation}
F(z) = \sum_{n=2}^{\infty} 
a_n z^{n-1} / (n-1)! 
\end{equation}
We find :
\begin{equation}
F(z) \sim \frac {\pi }{sin (\pi z)} - \frac{1}{z}
\end{equation}
If the Borel transformed function $F(z)$ would have no singularities
on the positive real axis, one could construct a regular
function $G(\Delta)$ having the correct perturbation expansion.
One takes the exponentially damped integral:
\begin{equation}
G(\Delta) = \int_0^{\infty} exp(-y/\Delta) F(y) dy
\end{equation}
However in our case 
the Borel transform has an infinity of poles for positive
values of the coupling constant $\Delta$. This means that there is
no unambiguous way to resum the perturbative series, so that
non-perturbative  effects much be present. 
In order to get a meaningful result one has to deform the contour
in the Borel plane. The correct choice of the contour is
in general arbitrary and some physical insight is necessary.
Here the situation is very similar to the calculation of the
one loop effective Lagrangian in an external electric field
in QED. The contour is deformed so as to give an imaginary part
to the effective Lagrangian, which has the physical interpretation,
that the electric field is instable under decay into 
electron-positron pairs. In the approximation we used in the
calculation, the vacuum also ought to be instable, since in
the bubblesum only the term $H \vec \varphi^2$ contributes, which
corresponds to a potential that is unbounded from below.

In complete analogy we therefore define the resummed series
for $\delta \rho$ by:
\begin{equation}
\delta \rho = \int_0^{-i \infty} exp(-y/\Delta) F_{\rho}(y) dy
\end{equation}
This way one find an imaginary part :
\begin{equation}
Im(\rho) = \frac{3 g^2 tg^2(\theta_W)}{64\pi^2}\frac{-i\pi}
{exp(1/\Delta) + 1}
\end{equation}
The non-perturbative form of the contribution is manifest.
Actually one can evaluate the complete contribution to
$\delta \rho$:
\begin{eqnarray}
& &\delta \rho = \frac{-3 g^2 tg^2(\theta_W)}{64\pi^2} \nonumber \\
& &(\psi(1/2 - i/(2\pi\Delta)) + log(2\pi\Delta) + i\pi/2 )
\end{eqnarray}
Here the $\psi$-function is the logarithmic derivative
of the $\Gamma$-function.
For large width $\Delta \rightarrow \infty$
one has the following asymptotic
expansion:
\begin{eqnarray}
& &\delta \rho = \frac{-3 g^2 tg^2(\theta_W)}{64\pi^2} \nonumber \\
& &(\gamma_{Euler} + 2log(2) + log(2\pi\Delta) + i\pi/2)
\end{eqnarray}
This shows, that if the Higgs is light, but has a large width,
the width plays the same role as a large Higgs mass, as far as
low energy radiative corrections are concerned.
 
\subsection{Direct summation}
While the approach described above is mathematically acceptable,
it is not quite satisfactory from the physical point of view.
The presence of an imaginary part to $\delta \rho$ is clearly
an artefact of the approximation, namely the effective use of
a cubic potential. Within the full theory this term cannot
be there, as the potential is positive definite. Also the separate 
summation of the higher-loop graphs, does not emphasize the connection
between loop-graphs and corresponding tree graphs. A more naive
calculation would consist of summing the bubbles in the Higgs
propagator first and then substituting the resummed propagator
into the loop-graph.

If one does this, one can consider the correction to the 
$\rho$-parameter as an averaged correction due to
a density of Higgs-fields. The density is the Kall\'{e}n-Lehmann
mass density $\sigma(s)$ of the Higgs propagator. 
In formula-form:
\begin{equation}
\delta \rho = \int ds \, \sigma(s)\, \delta \rho(m_H^2=s)
\end{equation}
In this from the origin of the problem gets localized in the
fact that the resummed Higgs propagator contains
a tachyon pole $m_T^2=-s_0 m^2_H$ at:
\begin{equation}
s_0 + 1 + \Delta \, log(s_0)=0
\end{equation}
The residue at the pole is given by $-s_0/(\Delta + s_0)$.
The presence of a tachyon is unacceptable and one has to
find a prescription to deal with it. Simply subtracting
the pole part is reasonable, as it leads to a Kall\'{e}n-Lehmann
density $\sigma(s)$
that is physical in the sense that $\sigma$ has support only for positive
s. However it leads to an undefined value of  the $\rho$-parameter.
The reason for this is that the integral $\int ds\, \sigma(s)$
is not unity anymore. One has to correct for this effect. The simplest
way is to add a non-perturbative factor $(\Delta + s_0)/\Delta$
to the Higgs-propagator. This last correction factor
is of course somewhat arbitrary, as one needs only an overall
normalization of the integral of $\sigma$. This kind of
uncertainty appears to be unavoidable, since one is not able
to solve the theory non-perturbatively. As the effect is however
suppressed by a factor $exp(-1/\Delta)$ the effect is very small 
in the case of a normal small coupling. The maximum correction
factor we get here is in the case  $\Delta = 1$ and is about
$28\%$. A graph
of the correction factor as a function of $\Delta$ is 
given in figure 1.

\begin{figure}[ht]
\centerline{\epsfig{file=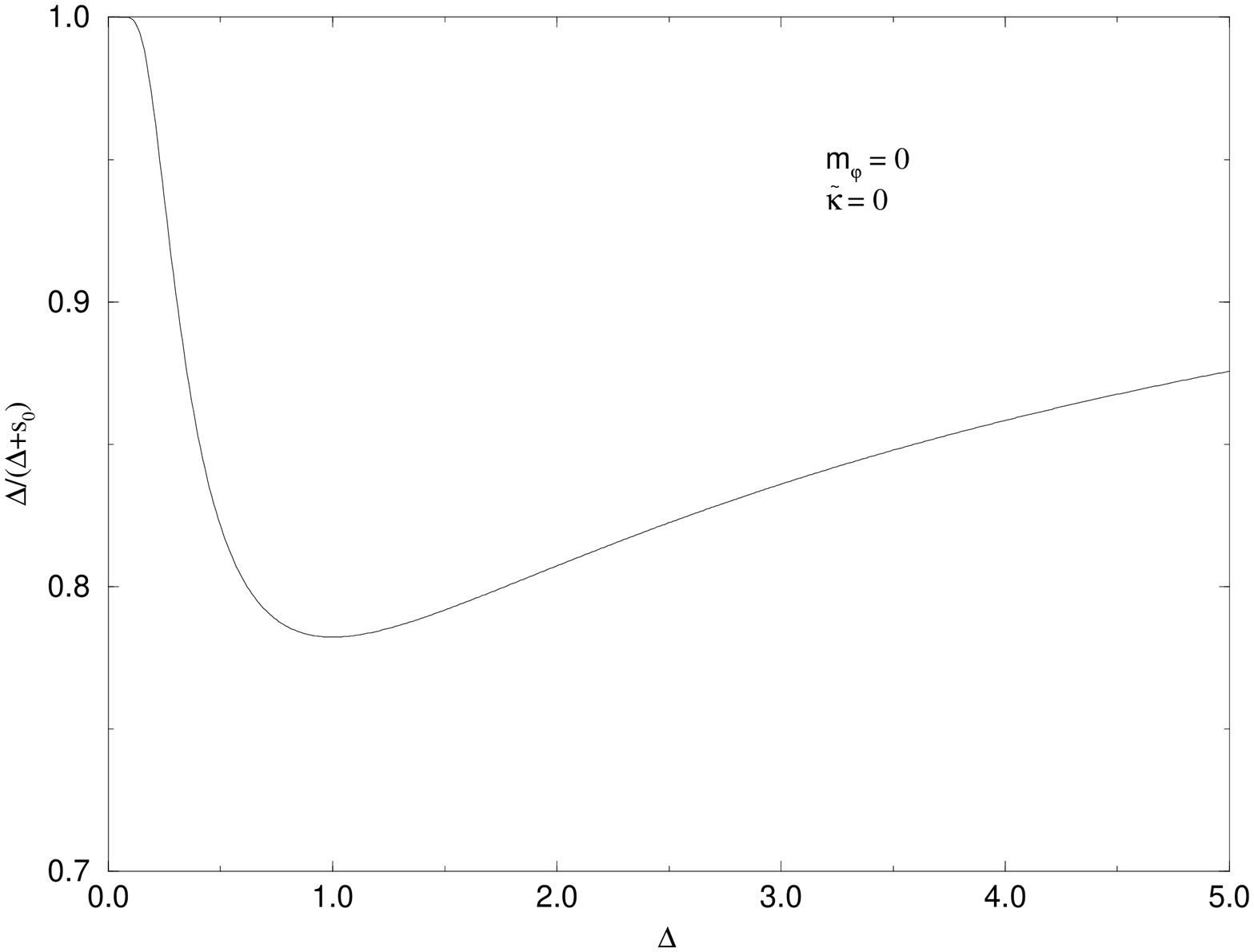,scale=0.37,angle=270}}
\caption{\label{fig1}
Non-perturbative propagator correction factor.}
\end{figure}

\section{CONCLUSION}
In conclusion, we have discussed the problem of
an unstable Higgs particle inside loops, both by a direct summation
and by substituting a resummed propagator in the loop.
In both cases an uncertainty in the result is present, due to
the fact that we cannot solve the theory non-perturbatively.
The uncertainty was quantified to be maximally around $28\%$.
Therefore qualitatively correct results can still be derived even
for a wide Higgs. In this case the radiative effects cannot be 
distinguished from the effects of a heavy Higgs particle.
The width starts playing the role of the Higss mass, which is easy
to understand on the basis of the Kall\'{e}n-Lehmann representation.
We  therefore conclude, that this model and similar ones with
a wide mass density cannot explain the dicrepancies in the
precision data, which were the original motivation of this study.
The model is not better, but also not worse than the
standard model.

While the method appears to work well in the present simple situation,
a number of open questions remain.
Foremost is the question, whether this method can be applied 
to gauge-theories.
Furthermore one should consider the possibility of going to higher 
orders in the $1/N$ expansion. Some work along these lines, 
within the heavy-Higgs standard model,
has been performed in \cite{ghin1}\cite{ghin2}. Finally
 connections with fully 
non-perturbative methods such as the lattice would be of interest.

\end{document}